# STATISTICAL ANALYSIS OF TOTAL COLUMN OZONE OVER UTTARAKHAND: ENVIRONMENT OF HIMALAYA

Namrata Deyal, Vipin Tiwari and Nandan S. Bisht*

Department of Physics Kumaun University SSJ Campus Almora-263601, Uttarakhand, India

*bisht.nandan@kunainital.ac.in

**ABSTRACT**

Total Column Ozone (TCO) is a critical factor affecting the earth's atmosphere, especially in the Himalayan region. A comprehensive study of TCO trend analysis and corresponding consequences in the Himalayan atmosphere needs to be demonstrated. We statistically examine TCO variability by analyzing the daily TCO dataset of the last 15 years (2005-2019) over the crucial region of the Himalayan environment i.e. Uttarakhand, India. Annual and seasonal trends of TCO variation have been analyzed and estimated by using robust statistical techniques i.e. linear regression method, Mann-Kendall test. Air mass trajectories have been estimated using Hybrid Single-Particle Lagrangian Integrated Trajectory (HYSPLIT) to understand the source of air pollutants and corresponding continental and maritime transportation towards Uttarakhand under various climatic conditions. Obtained results indicate that TCO values are at peak during the spring season whereas it shows the least value during the winter season over Uttarakhand. The highest and lowest value of Coefficient of Relative Variance (CRV) is estimated as 3.14 and 1.09 during winter and monsoon season respectively. Moreover, Least Square Method (LSM) and the Mann–Kendall test estimate a high correlation (0.86) for the seasonal and annual trend of TCO. Further results indicate that the inter-annual oscillation pattern of TCO is similar to Quasi-Biennial Oscillation (QBO) significantly. In addition, a comparative study has been performed for the data measured by two TCO measuring instruments i.e. Ozone Monitoring Instrument (OMI) and Ozone Mapping Profiler Suite (OMPS). TCO values measured from both instruments are highly correlated (0.96%) with an average relative difference of around 3%. The outcomes of this study are expected to be beneficial for future study of TCO over other crucial regions of Himalayan territory.

# INTRODUCTION

Ozone is a robust component of the earth's atmosphere and can be found primarily in two layers of earth atmosphere. A major part i.e. 90% of its mass is found in the stratosphere lying within 16-35 kilometers above the surface of the earth whereas the remaining 10% ozone is found in the troposphere of earth. Stratospheric ozone is considered as shielding layer (good ozone) as it hinders the high energy radiations (2000 $^0$A to 3000 $^0$A) from penetrating the earth's main environment. On the other hand, Ultra Violet (UV)-B radiations are biologically harmful solar radiation and the ozone layer prevents these hazardous radiations to reach on the surface of the earth (Dobson et al., 1973; Dobson et al., 1968; Madronich et al., 1998). Most ozone is formed in the tropic region where it moves from lower to higher latitude by the Brewer-Dobson circulation (Weber et al., 2011).TCO is the amount of total ozone (troposphere and stratospheric ozone) containing in a vertical column from the earth surface to the top of the atmosphere at standard temperature and pressure (STP) (David W et al., 2014; Brasseur et al., 1988 Brasseur et al., 1997). It is measured using data obtained from ground-based stations and satellites in terms of Dobson unit (DU), which describes the thickness of a layer of pure ozone at STP. One DU is $2.69 \times 10^{16}$ ozone molecules/cm$^2$ in a vertical column extended till the atmospheric limit or 0.01mm thick layer of pure ozone to the surface of the earth at standard condition (STP). The value of TCO significantly varies from 200 to 300 DU all over the globe. Meanwhile, if its value reduces to 220 DU it is considered as a condition for occurring Ozone hole at any location.

A series of reactions of ozone formation and distortion was proposed by Sydney Chapman in 1930. It depicts that the photolysis of Oxygen molecules produces nascent radical in presence of UV radiation which further reacts with oxygen and results in ozone molecule. Moreover, ozone depletion involves the reverse process of conversion of ozone molecules into atomic oxygen due to catalyst reaction [Groves et al., 1980]. Apart from this ozone abundance are the key parameters for estimating the formation and depletion of ozone at any location of the globe. The processes of formation and deterioration of ozone are found to be naturogenic as well as anthropogenic. However, both can occur simultaneously (Rubin et al., 2001). The anthropogenic chemicals such as chlorofluorocarbons (CFCs), halocarbons, and other volatile organic compounds can destroy stratospheric ozone significantly by reacting with ozone. (Solomon, S. et al., 1999; Cicerone et al., 1974; Crutzen et al., 1974). Apparently, few studies suggest long term variability of ozone with geophysical parameters, including the Quasi-Biennial Oscillation, solar activity, the stratosphere-troposphere exchange, and volcano eruption (Willett, H.C. et al., 1962; Garcia, R. et al., 1987; Ningombam S S, et al., 2011; Hema Bisht et al., 2014).

Despite a small portion of ozone contributes to earth' atmospheric composition, TCO plays a vital role in climate change and ecology (Wayne et al., 1987; Wayne et al., 2000; Ogunjobi et al., 2007). The depletion of the ozone layer results in severe serious health issues i.e. skin cancer, cataract, deficient immunity, respiratory disorders for species (UNEP 1998; UNEP 2003; UNEP 2016). It can also contaminate the marine ecosystem as well (cell-damaging life of phytoplankton and zooplankton). Besides, ozone depletion permits UV-B radiation which may retard the physiological and developmental processes of plants (Schmalwieser et al., 2003; Sivasakthivel et al., 2011). Therefore, ozone layer conservation is mandatory to sustain an ideal ecological surrounding on earth (WMO 1989; WMO 2011, J.I Freijer et al., 2002; R. Venkanna et al., 2015;).

Numerous studies on atmospheric ozone have been carried out all across the world. The Antarctic ozone hole was first observed by the British Antarctic Survey in the 1980s (Farman et al., 1985). Several studies emphasize seasonal depletion in the stratospheric ozone layer in Antarctica and have observed depletion during the spring season (Iwasaka et al., 1987; Rowland et al., 1989; Stolarski et al., 1990). Over the last two decades, minute depletion in ozone was also observed over the Arctic during the late winter and early spring (Brune et al., 1990 Goutail, F., et al., 1999). But these studies are limited to the polar region only (Newman et al., 2006). Such studies have been extended for other locations of the globe and established statistical trends of TCO over the northern and southern mid-latitudes (Rowland et al., 1991; Chakrabarty et al., 1998; By Michael L. et al., 2007; P. J. Nair et al., 2013; Ayodeji Oluleye et al., 2013; Kok Chooi Tan et al., 2014; A. Badawy et al., 2017). These studies suggest that the depletion of ozone is not only confined in the Polar Regions but vary significantly in another latitudinal region as well.

The study of ozone layer depletion and variation has emerged as a major global scientific and environmental issue in the recent past. In this framework, various international protocols have been designed (UNEP 1987). Montreal Protocol is an international treaty to protect the ozone layer by phasing out the production and limited use of ozone-depleting substances (ODS) (WMO 1989). This agreement was executed on 1Jan 1989 and has been signed by most of the countries including India. This protocol has reduced the risk of further ozone depletion to a large extent (Zubov et al., 2001; Anderson et al., 2000; Guus J. M. Velders et al.,2007; Pinedo Vega et al., 2017). From the perspective of India, a few studies on ozone variation and depletion have been reported to date (Mani et al., 1973; Kundu et al., 1993; Pulikesi et al., 2006; Vazhathottathil Madhu et al., 2014). These studies are limited to industrial areas with high population density. However, there is a paucity of such studies in Himalayan regions i.e. areas with lower population density and enhanced ecological versatility. The ozone climatology is supposed to be crucial in these regions due to climatic disparity. Therefore, a comprehensive study of ozone (TCO variation) over the Himalayan region is needed to monitor the atmospheric ozone concentration in a regular period.

In this paper, we have statistically analyzed the spatial and temporal tend of TCO variability using the TCO dataset of the last 15 years i.e. 2005-2019 over the crucial region of the Himalayan environment i.e. Uttarakhand, India. Air mass trajectories have been estimated using Hybrid Single-Particle Lagrangian Integrated Trajectory (HYSPLIT) to understand the source of air pollutants and corresponding continental and maritime transportation towards Uttarakhand under various climatic conditions has been investigated. We have compared the data measured by two TCO measuring instruments i.e. Ozone Monitoring Instrument (OMI) and Ozone Mapping Profiler Suite (OMPS) and corresponding results have been analyzed.

## 2. METHODOLOGY

### 2.1 Study Area

Uttarakhand, a state in northern India crossed by the Himalaya extended from ($28.71\,^0N$ to $31.45\,^0N$ and $77.56\,^0E$ to $81.03\,^0E$) is the study area shown in Fig. 1. It is situated in southern Asia (North of Indo- Gangetic Plain (IGP), East of China, and foothill of Himalayas). It has a total geographic area of $53,483\ km^2$ of which 86% is mountainous and 65% is covered by forest with an average population density of 189 people/$km^2$ (Census 2011). It is also surrounded by major

Indian industrialized cities i.e. Delhi, Lucknow, and Chandigarh. Along with more industrial development the agricultural activities, burning of crop residues, and forest fire are the primary source of pollutants over Uttarakhand. The mountainous topography and complex land-sea interactions are responsible for phenomenal changes in weather over Uttarakhand. The average temperature varies within a range of 21.8 $^0$C to 5 $^0$C with the highest in June while the lowest in December and for few stations the temperature falls below 0 $^0$C. Maximum rainfall occurs from July to September i.e. early NEM with average precipitation 160-200 cm. Uttarakhand holds a rich heritage of India and is known as Devbhumi i.e. "Land Of God". Its snow-clad peaks, beautiful hill stations, and charming weather make it a favorite destination for tourism. Nanda Devi, Valley of flowers and National Park (1988, 2005) are world heritage sites declared by UNESCO are the most fascinating places of Uttarakhand. Uttarakhand is the most popular Himalayan state of India, situated in the natural environment of Himalaya and it plays a vital role in hosting many animals, plants, and rare herbs. Most importantly, Uttarakhand covers a major portion of the Himalayan range and therefore represents a crucial study station (area) to analyze atmospheric phenomena i.e. ozone variation over the Himalayan region.

**2.2 Data Sources**

The daily data of TCO has been acquired from the open resource database of NASA and NOAA. The used data is recorded with two TCO measuring instruments i.e. Ozone Monitoring Instrument (OMI) and Ozone Mapping and Profiler Suite (OMPS). The TCO value is measured in Dobson Unit (DU). The details about the ozone measurement are summarized in Table 1.

The monthly and annual mean for all the stations are derived by taking the cumulative mean of daily data. The principle of measurement of TCO is based on the reflection of solar radiation on certain spectrum bands (Heath et al., 1975; P.K.Bhartia et al., 2013), some unresolved data corresponding to polar night and low light area cannot be measured and recorded as zero (Pinedo-Vega et al., 2017). These unresolved data are few and removed in to account for monthly and annual mean. To examine the seasonal variability of ozone we divide TCO data into five seasons' i.e. winter (Dec-Feb), spring (March-April), summer (May-June), monsoon (July-Sep), and autumn (Oct-Nov). Also, to investigate the variation in ozone profile i.e. the latitudinal variation of TCO, we studied four stations of India i.e. Leh Laddak (34.10N-77.4E), Dehradun (28.37N-77.13E), Banglore (12.58N-77.34E), and Kanyakumari (8.4N-77.32E) from northward to southward for the year 2014. The coefficient of Relative Variance (CRV), correlation coefficient, probability distribution, least square method i.e. regression analysis method, and Mann–Kendall test (MK) are used for statistical analysis and trend analysis of TCO variation over Uttarakhand during a period of 15 years i.e. from 2005 to 2019.

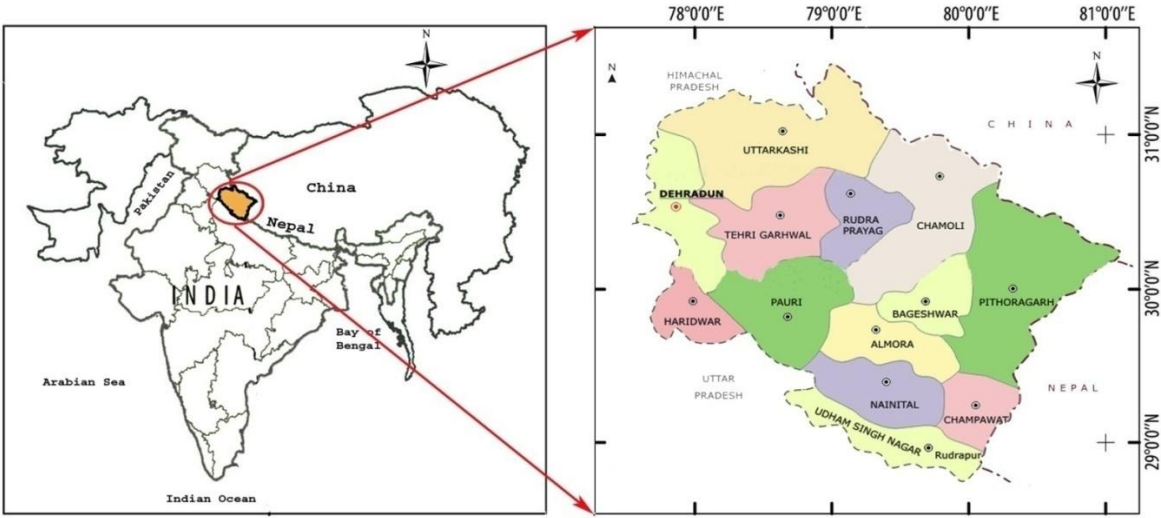

Figure 1: Map of Uttarakhand with latitude and longitude of the study area.

Table 1: Locations and description of the selected station of Uttarakhand

| DISTRICT | ALTITUDE($^0$) | LONGITUDE($^0$) |
|---|---|---|
| U.S NAGAR | 28.961N | 79.515E |
| PITHORAGHARH | 29.582N | 80.218E |
| ALMORA | 29.589N | 79.646E |
| HARIDWAR | 29.945N | 78.164E |
| DEHRADUN | 30.316N | 78.032E |
| CHAMOLI | 30.403N | 79.336E |

Table 2: Characteristics of the total ozone content datasets.

| Satellite | Aura spacecraft | Suomi NPP & NOAA-20 |
|---|---|---|
| Instrument | Ozone Monitoring Instrument (OMI) | Ozone Mapping And Profiler Suite (OMPS) |
| Parameter | TCO | TCO |
| Data Type | Daily | Daily |
| Period | Jan- 2005 to Dec-2019 | Jan- 2012 to Dec-2019 |
| Continue | Total Ozone Mapping Spectrometer (TOMS) series data | Solar Backscatter Ultraviolet (SBUV) series data |
| Spatial Coverage | Global ($0.25^0$ log * $0.25^0$ lat) | Global ($1^0$ logs * $1^0$ lats) |

## 2.2 METHODS

a) Percentage relative difference is calculated as;

$$\% \text{ difference} = \frac{TCO_{(OMI)} - TCO_{(OMPS)}}{TCO_{(OMPS)}} \quad (1)$$

Here 'X' is the value of ozone concentration and N is the number of years.

b) A coefficient of relative variation (CRV) for each site has been determined as;

$$CRV = \frac{SD}{\mu} \times 100 \quad (2)$$

Here 'SD' is the standard deviation and '$\mu$' is the mean for N years.

c) Trend analysis and fluctuation of ozone concentration have been estimated by Linear Regression (Least Square method) and Mann-Kendall test for all the selective sites.
d) Statistical models are used for the probability distribution (Normal, Lognormal, and Gamma) of TCO. We check their deviation from normality using KST (Kolmogorov–Smirnov D-test) probability distribution over Uttarakhand.

## 2.3 Back Trajectory Calculation

The origin of air quality and transportation patterns of air mass arriving towards Uttarakhand has been analyzed using Hybrid Single-Particle Lagrangian Integrated Trajectory (HYSPLIT) backward trajectories. The HYSPLIT Model (Draxler and Rolph, 2011) developed by Air Resource Laboratory (NOAA) and these trajectories have been simulated from the Web site (http://ready.arl.noaa.gov/HYSPLIT_traj.php). Based on the highest concentrations of ozone in the study period, multiple air mass back trajectories for 14th May 2014, 13th June 2014, and 22nd Dec 2014 are stimulated at different elevations: 500 and 1,000 m a.g.l. using HYSPLIT models. The analysis has been performed with the GDAS meteorological dataset at a starting time of 00:00 UTC with a total run time of 120 hours.

## 3 RESULTS & DISCUSSION

The daily data of TCO (DU) from January 2012 until December 2019 has been acquired from OMI and OMPS. The monthly mean of TCO from both the datasets has been compared by estimating relative differences based on Equation (1) have been tabulated in Table 3. A range of relative difference 0.51% to 5.29% from 2012 to 2019 has been found which is within the acceptable interval. Moreover, the average relative difference of monthly TCO data via both the datasets is merely 3.5%. Also, we found a high positive correlation between the values of TCO retrieved from both the instruments having a different resolution for the period of 2012-2019.

We have not observed an average monthly mean of TCO value less than 220 DU (Condition for occurring Ozone hole at any location on the globe). It implies that However, we noted TCO value less then 220DU for one or two days in December 2008 and 2016 and January 2019 from the OMI data set while in December 2008 and January 2019 from the OMPS data set, which can be ignored for accounting it as the ozone hole. For the total number of years considered average ozone maximum and minimum value have observed 395 DU and 212 DU from OMI and 382DU and 208 DU from OMOPS respectively.

Table 3: Comparison of TCO values retrieved from OMI and OMPS datasets.

| | Year/Month | 2012 | 2013 | 2014 | 2015 | 2016 | 2017 | 2018 | 2019 |
|---|---|---|---|---|---|---|---|---|---|
| | JAN | 0.88 | 8.25 | 2.24 | 1.35 | 3.56 | 2.65 | 0.12 | 1.06 |
| | FEB | 1.16 | 0.60 | 0.14 | 0.34 | 4.59 | 6.40 | 0.89 | 2.90 |
| | MARCH | 1.81 | 4.80 | 2.49 | 5.87 | 1.55 | 3.51 | 1.75 | 5.52 |
| % difference between OMI and OMPS data of TCO | APRIL | 10.11 | 0.02 | 1.01 | 1.57 | 4.06 | 5.32 | 1.05 | 0.17 |
| | MAY | 4.28 | 10.47 | 10.45 | 5.71 | 9.47 | 5.41 | 1.23 | 0.98 |
| | JUNE | 5.13 | 5.92 | 7.02 | 6.32 | 8.20 | 5.24 | 0.31 | 0.17 |
| | JULY | 9.17 | 10.43 | 8.00 | 7.40 | 10.18 | 4.42 | 0.13 | 0.46 |
| | AUG | 7.59 | 8.10 | 5.76 | 5.33 | 6.75 | 5.58 | 0.09 | 0.23 |
| | SEP | 4.10 | 4.98 | 2.30 | 1.14 | 2.16 | 2.59 | 0.02 | 0.85 |
| | OCT | 1.62 | 3.69 | 4.41 | 0.03 | 1.39 | 2.09 | 0.32 | 0.25 |
| | NOV | 2.77 | 3.79 | 5.03 | 3.24 | 4.59 | 5.29 | 0.00 | 1.86 |
| | DEC | 1.02 | 2.44 | 0.99 | 7.70 | 4.64 | 1.47 | 0.26 | 1.06 |
| | AVERAGE | 4.14 | 5.29 | 4.15 | 3.83 | 5.10 | 4.16 | 0.51 | 1.29 |
| The correlation coefficient between the value of TCO with OMI and OMPS data | | 0.86 | 0.69 | 0.68 | 0.73 | 0.85 | 0.86 | 0.99 | 0.96 |

### 3.1 Monthly Variation

The monthly mean of TCO derived from the cumulative mean of daily data of TCO is found in the range from (297± 13.78) DU to (260± 8.73) DU for OMI and (300 ±16.08) DU to (258±6.07) DU from OMPS datasets. The average maximum and minimum value of ozone concentration are measured in May and December respectively from both the instruments over Uttarakhand. Further, a bell-shaped normal distributed curve for monthly ozone mean of 15 years (2005-2019) has been observed as shown in Figure 4. We have observed similar results as the average TCO value from OMI increased, the ozone value measured from OMPS was also increased. Therefore, the quality of OMPS datasets is analogous to OMI data of ozone.

We have statistically analyzed the data and estimated final distributions comply by the mean monthly series of TCO by applying Kolmogorov–Smirnov goodness of fit test of distribution. It is noticed that monthly TCO distribution exhibits lognormal probability distribution in Uttarakhand. The basic statistics (maximum, minimum, average, standard deviation, and distributions) for the Interannual monthly average TCO (2005-2019) from both data sets are mentioned in Tables 3 and 4. On observing table 3 and 4, we have found that the set of monthly series of TCO follows different distributions and changes from normal, lognormal to gamma distribution which indicates that the distribution is progressively approaching heavier tailed from January to December.

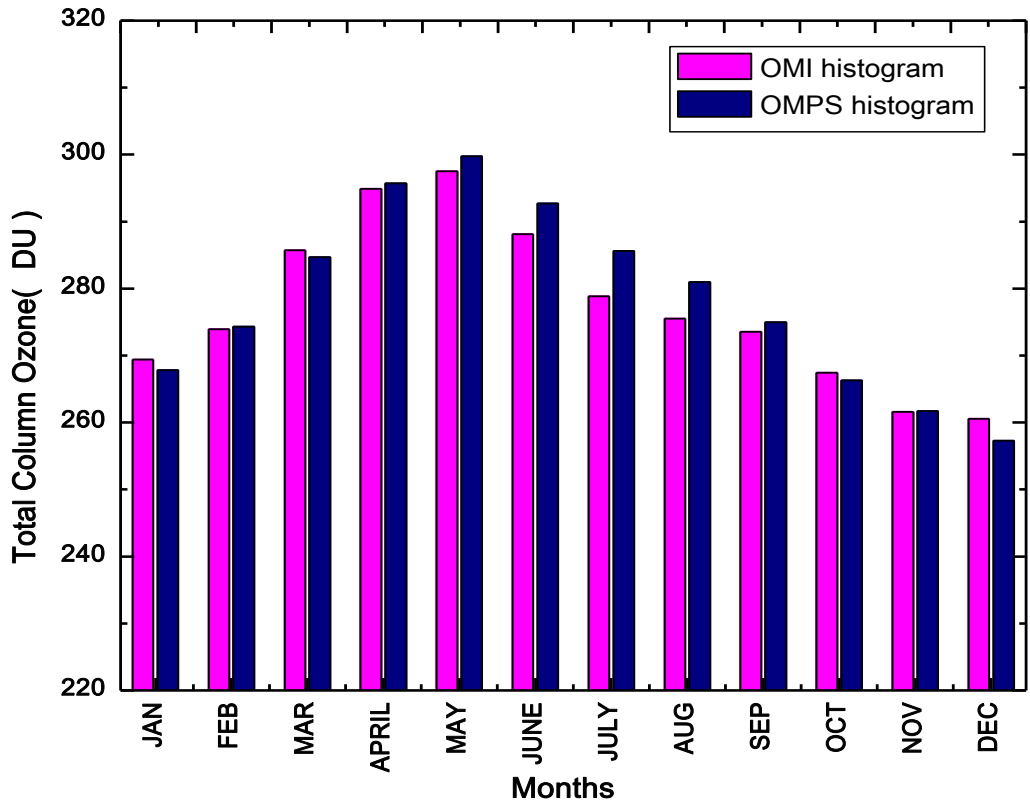

Figure 2: Monthly variation of TCO over Uttarakhand via OMI and OMPS datasets.

**Table 4: Probability distribution with statistical parameters of TCO via OMI datasets (2005–2019) for Uttarakhand**

| Months | Max. | Min. | Mean value | Standard deviation | Distribution | KST-value α= 0.05 |
|---|---|---|---|---|---|---|
| January | 279.24 | 250.97 | 269.39 | 9.25 | Normal | 0.1964 |
| February | 298.31 | 254.78 | 273.95 | 13.70 | Lognormal | 0.1891 |
| March | 305.75 | 268.71 | 285.74 | 8.94 | Lognormal | 0.1315 |
| April | 313.14 | 282.11 | 294.91 | 10.33 | Lognormal | 0.1453 |
| May | 319.07 | 278.42 | 297.51 | 13.78 | Gamma | 0.1692 |
| June | 312.28 | 275.72 | 288.12 | 11.33 | Lognormal | 0.2054 |
| July | 301.93 | 264.86 | 278.86 | 12.25 | Lognormal | 0.1890 |
| August | 294.97 | 266.14 | 275.53 | 10.44 | Lognormal | 0.1812 |
| September | 287.97 | 265.80 | 273.53 | 6.74 | Lognormal | 0.1760 |
| October | 278.43 | 258.38 | 267.44 | 6.34 | Lognormal | 0.2503 |
| November | 275.72 | 249.34 | 261.62 | 8.10 | Gamma | 0.1789 |
| December | 273.30 | 241.25 | 260.58 | 8.73 | Normal | 0.1218 |

**Table 5: Probability distribution with statistical parameters of TCO via OMPS datasets (2005–2019) for Uttarakhand**

| Months | Max. | Min. | Mean value | Standard deviation | Distribution | KST-value α= 0.05 |
|---|---|---|---|---|---|---|
| January | 290.60 | 245.48 | 266.77 | 11.47 | Normal | 0.1762 |
| February | 297.63 | 242.39 | 273.80 | 15.47 | Lognormal | 0.1577 |
| March | 296.32 | 274.06 | 285.21 | 6.12 | Lognormal | 0.1034 |
| April | 318.56 | 281.81 | 295.67 | 10.71 | Gamma | 0.1827 |
| May | 321.58 | 274.14 | 300.12 | 16.08 | Normal | 0.1634 |
| June | 308.96 | 275.25 | 293.49 | 11.40 | Lognormal | 0.1554 |
| July | 301.38 | 267.77 | 286.47 | 12.44 | Lognormal | 0.2055 |
| August | 295.00 | 268.93 | 281.73 | 9.66 | Lognormal | 0.1841 |
| September | 287.15 | 266.63 | 275.57 | 5.79 | Gamma | 0.1360 |
| October | 277.77 | 257.18 | 266.83 | 6.08 | Lognormal | 0.1647 |
| November | 276.59 | 250.00 | 262.36 | 7.57 | Gamma | 0.1283 |
| December | 267.81 | 247.97 | 258.10 | 6.07 | Gamma | 0.1389 |

### 3.2 Inter Annual Variation and Seasonal Variation

Box plots of average annual and seasonal values of TCO (2005-2019) from both the datasets were illustrated in figure 3. It is observed that the summer and autumn seasons have the highest and lowest average TCO value from both the sensors. Seasonal average time series shows that average TCO in the summer season was maximum (292.81± 11.82) DU and minimum in autumn (264.53 ± 6.87) DU from OMI dataset whereas OMPS shows that the maximum average TOC in the spring season was (297.73± 12.16) DU and the minimum was (259.48 ±6.58) autumn DU. In addition, the sensors recorded annual maximum and minimum TCO measurements in the same years 2015 and 2008 respectively. The behavior is not repeated in any of the seasons (spring, summer, monsoon, and autumn) except winters i.e. the maximum and minimum TCO values from two sensors do not overlap in the same year again.

Interannual variability of TCO retrieved from the OMI dataset (2005 – 2019) showed merely the same oscillating behavior for all the stations along the latitudinal and longitudinal belt of Uttarakhand illustrated in Figure 4. From 2005 to 2019 ozone variability repeats its ascending and descending pattern at a period of two years. This fluctuation of TCO follows the analogous pattern of the quasi-biennial oscillation (QBO) significantly. Further, we observed an abrupt change in TCO in 2007 and 2016. It suggests this fluctuation might be due to interruption in QBO for the respective years.

Table 4 depicts the trend of TCO annual and seasonal (retrieved from OMI from 2005 to 2019) for all the stations by the least square method. We obtained a significant negative trend of TCO with rate 0.03 DU/year, 0.162 DU/year, -0.045DU/year, -0.057DU/year, -0.001 DU/year, and 0.033DU/year for Dehradun, Almora, Haridwar, Pithoragarh, U.S.Nagar and Chamoli respectively. However, the annual decreasing rate is very low over Uttarakhand which indicates a significant recovery of TCO during the study period. Also, we observed a negative trend for winters and spring seasons for most of the stations while the increasing trend in summers, monsoon, and autumn seasons for all the stations. Similarly, results obtained from the Mann–Kendall test are illustrated in table 5. The trend by LSM and MK test is highly correlated in all the season as well as annually except monsoon. The correlation coefficient between trend obtained from both the methods has been calculated 0.86 annually, 0.72 in winters, 0.97 in spring, 0.53 in summer, 0.24 in monsoon, and 0.75 in autumn seasons respectively. Moreover, for most of the year, the value of TCO has been observed to fluctuate about its mean value excluding in 2008 and 2015.

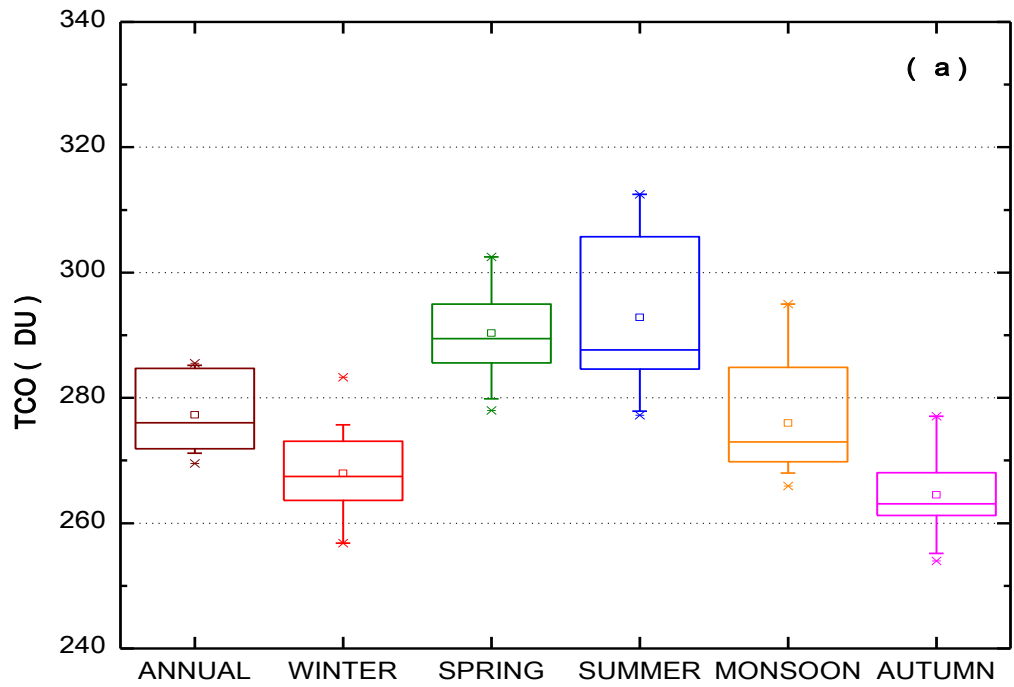

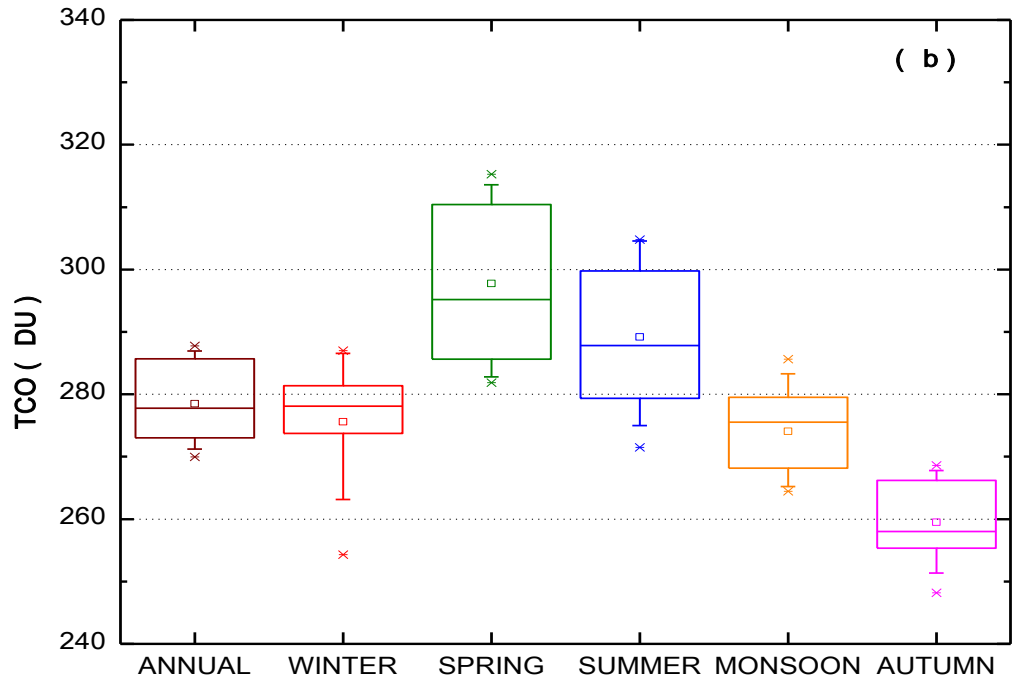

Figure 3: and the seasonal average of TOC (2005–2019) from (a) OMI and (b) OMPS datasets

Table 6: Annual and seasonal trend of TCO by least square method from 2005-2019

| LSM | annual | winter | spring | summer | monsoon | autumn |
|---|---|---|---|---|---|---|
| Dehradun | 0.03 | -0.589 | 0.09 | 0.078 | 0.084 | 0.051 |
| Almora | 0.162 | -0.144 | 0.226 | 0.231 | 0.146 | 0.285 |
| Haridwar | -0.045 | -0.495 | -0.041 | 0.69 | 0.191 | 0.018 |
| Pithoragarh | -0.057 | -0.574 | -0.114 | 0.013 | 0.095 | 0.078 |
| U.S. Nagar | -0.001 | -0.579 | -0.459 | 0.14 | 0.027 | 0.159 |
| Chamoli | 0.033 | -0.489 | 0.32 | 0.058 | 0.089 | 0.056 |

Table 7: Annual and seasonal trend by Mann-Kendall test from 2005-2019

| MK TEST | annual | winter | spring | summer | monsoon | autumn |
|---|---|---|---|---|---|---|
| Dehradun | 0.30 | -0.49 | 0.00 | 0.59 | 0.59 | 0.49 |
| Almora | 0.60 | -0.25 | 0.40 | 0.60 | 0.40 | 0.79 |
| Haridwar | 0.20 | -0.30 | 0.00 | 0.79 | 0.69 | 0.20 |
| Pithoragarh | 0.10 | -0.49 | -0.30 | 0.59 | 0.40 | 0.49 |
| U.S. Nagar | 0.00 | -0.69 | -1.44 | 0.25 | 0.55 | 1.04 |
| Chamoli | 0.30 | -0.49 | 0.49 | 0.59 | 0.10 | 0.40 |

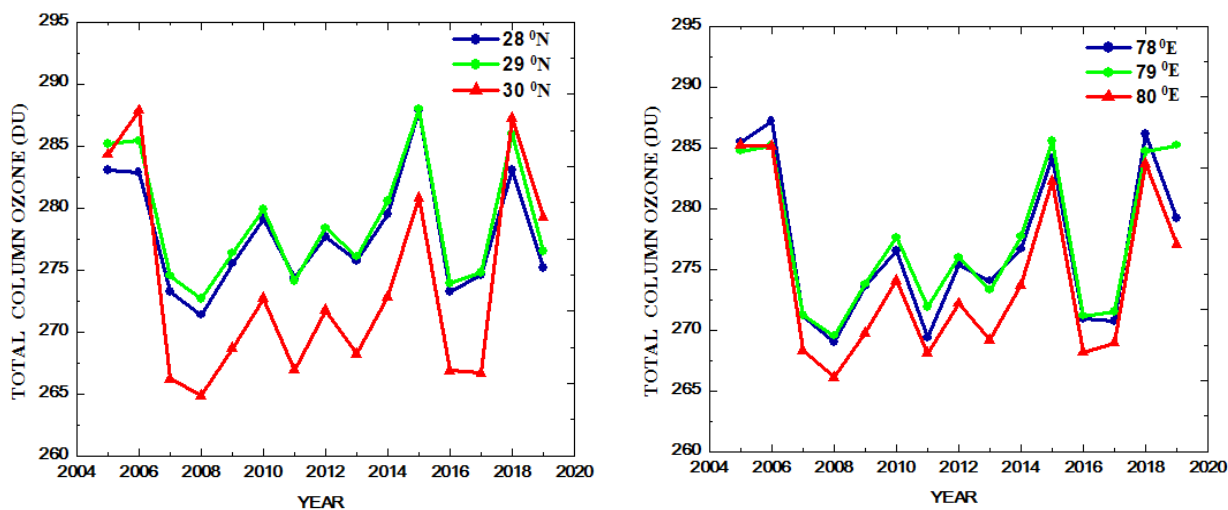

Figure 5: The Interannual variation of TCO across (a) longitudinal and (b) latitudinal belt of Uttarakhand

## 3.3 Spatial Variation (Coefficient of relative variation (CRV))

The variability of ozone concentration over Uttarakhand has been studied. Annual and seasonal CRV of TCO for six stations i.e. Dehradun, Almora, Haridwar, Pithoragarh, U S Nagar, and Chamoli is shown in Table 5. A significant variation in TCO has been observed among these study areas. Further, it is clear from the observations that the highest value of CRV is obtained in winter while the lowest at monsoon for all the selective stations. However, a gradual ascending trend has been observed for the annual variation of CRV from southern stations to northern stations. These outcomes suggest that annual TCO varies as a function of latitude. On observing these trends, it is a noticeable fact that CRV of ozone concentration decreases gradually from north to south in winter, whereas it increases from west to east in summer and autumn. Also, we found that in monsoon season it decreases from west to east. To have more clear insight, the spatial distribution of CRV is shown in fig. 5.

Table 8: Coefficient of the relative variance of TCO from 2005-2019.

| Station | Winter | Spring | Summer | Monsoon | Autumn | Annual |
|---|---|---|---|---|---|---|
| Dehradun | 3.11 | 2.35 | 2.29 | 1.18 | 2.35 | 1.60 |
| Almora | 2.93 | 2.32 | 2.36 | 1.24 | 2.08 | 1.67 |
| Haridwar | 2.95 | 2.24 | 2.09 | 1.19 | 2.03 | 1.58 |
| Pithoragarh | 3.13 | 2.35 | 2.38 | 1.21 | 2.01 | 1.66 |
| U.S.Nagar | 2.77 | 2.58 | 2.11 | 1.24 | 1.85 | 1.64 |
| Chamoli | 3.14 | 2.68 | 2.42 | 1.09 | 2.37 | 1.70 |

## 3.4 Latitudinal Variation

To characterize the ozone profile i.e. latitudinal distribution of TCO, we have estimated the monthly mean of ozone concentration for four Indian subcontinent Leh Laddak (34.10N-77.4E), Dehradun (28.37N-77.13E), Banglore (12.58N-77.34E), and Kanyakumari (8.4N-77.32E) for the year 2014. Fig. 6 indicates the legging of two months in maximum ozone value for Leh (high latitudinal area) and Kanyakumari (low latitudinal area). We have statistically analyzed the data and estimated skewness for distribution for northward region i.e. Leh Laddak and Dehradun the value of skewness is found positive (0.515 and 0.183) respectively. Similarly, the skewness is negative i.e. -0.539 and -.600 for Banglore and Kanyakumari respectively. However, a quantitative variation in TCO approx 16% has been calculated from a latitudinal belt of India i.e. from Kanyakumari to Leh Ladakh. The probable reason for this variation might be the processes of formation and transportation ozone from the tropic to the pole region. (Chakrabarty et al., 1979; Nandita D. Ganguly et al., 2005). Also, CRV has been estimated at 6.58, 4.8, 4.48, and 4.17 for Leh Laddak, Dehradun, Banglore, and Kanyakumari respectively.

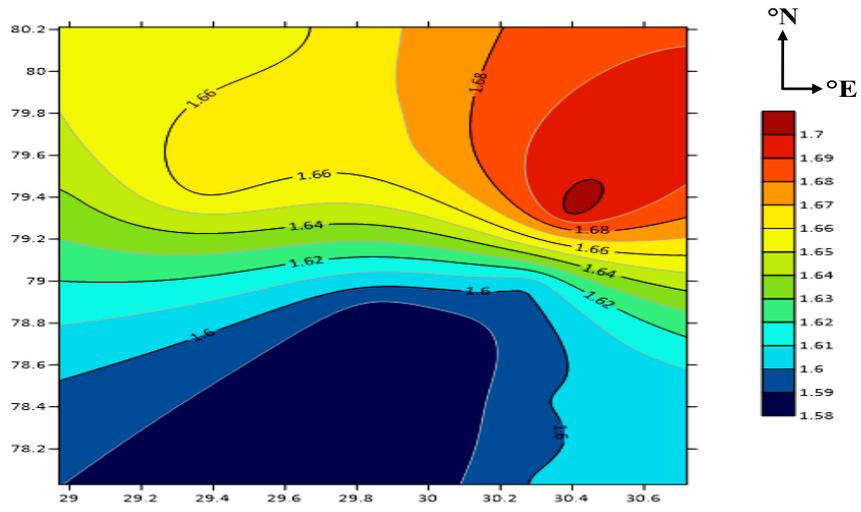

(a)

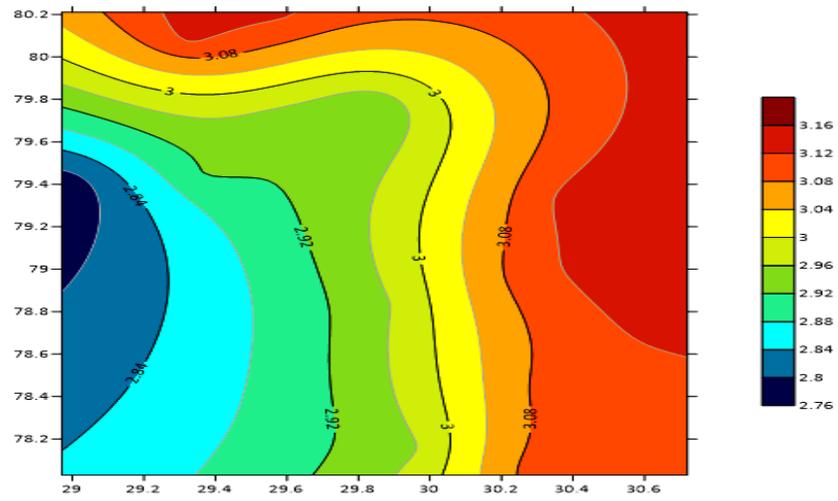

(b)

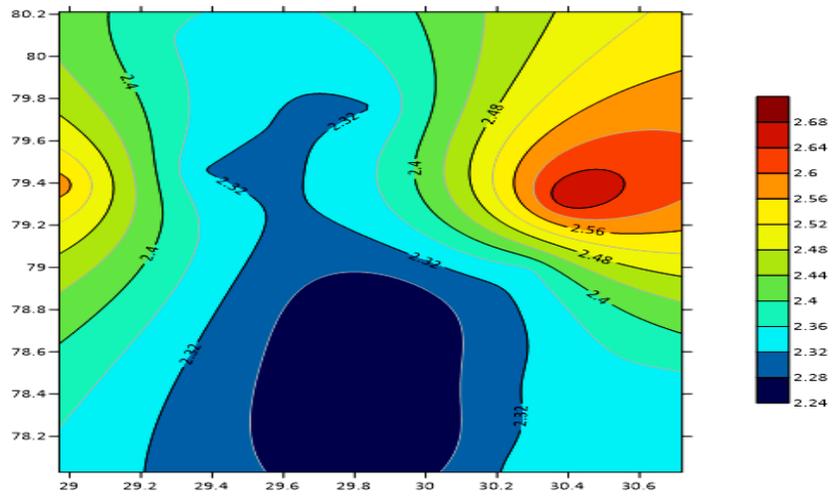

(c)

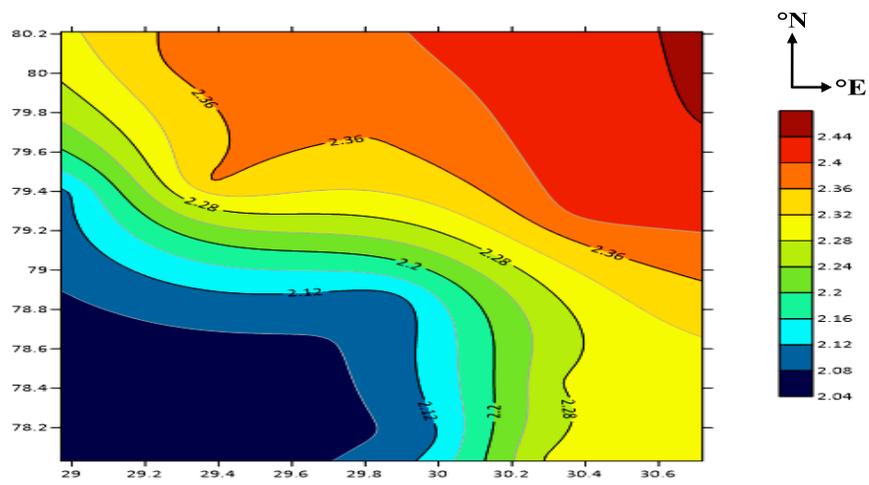

(d)

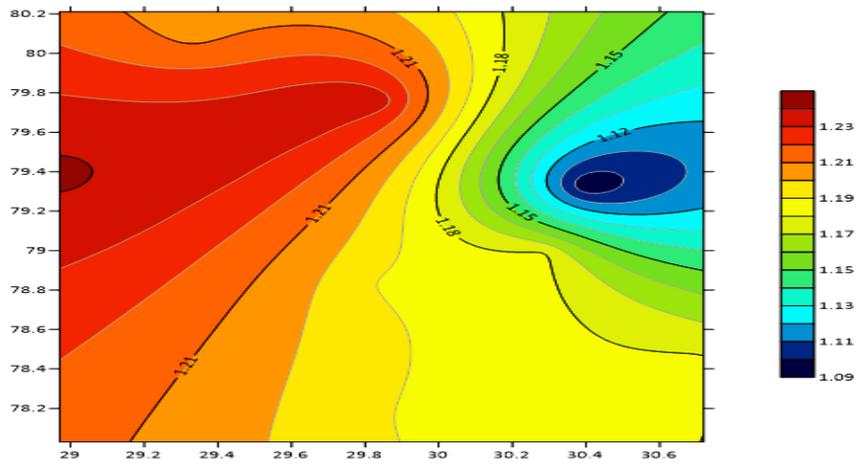

(e)

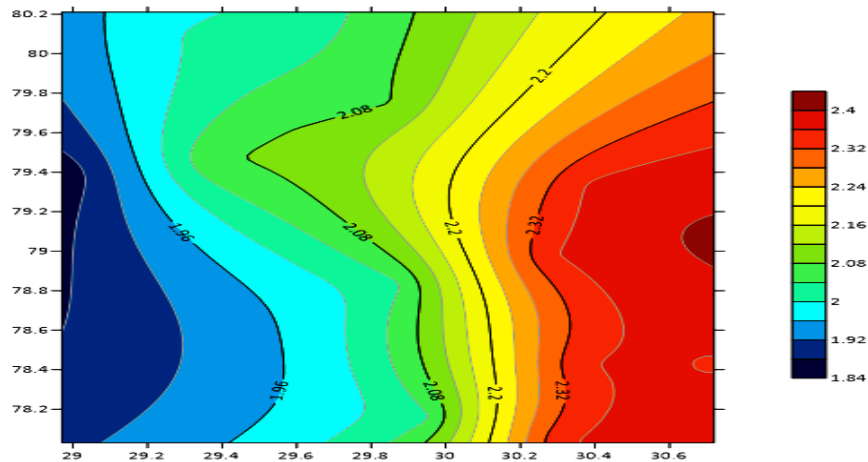

(f)

Figure 5: CRV distribution of ozone over Uttarakhand (a) Annual (b) Winter (c) Summer (d) Spring (e) monsoon (f) autumn

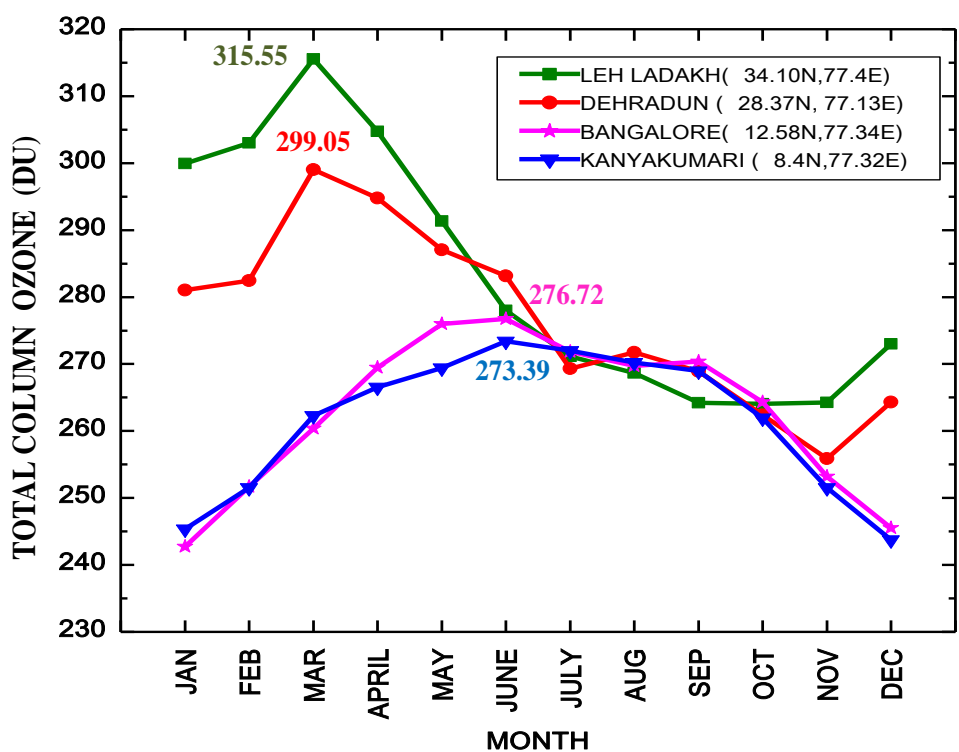

Figure 6: Latitudinal variation of TCO from southward to northward regions in the year 2014

### 3.5 Back Trajectory Analysis

Back trajectories are simulated for correlating it with seasonal change of TCO and to understand the air mass transport and source of origin (Stohl, A., 1998). Further observing the influence of continental and maritime air mass on seasonal variation of surface ozone, backward air trajectories are retrieved from the surface of Uttarakhand fig 7(a), 7(b), and 7(c) during seasons i.e. spring, summer, and winter. These back trajectories imply that the air mass moves from west of Uttarakhand summer and winter seasons. During the spring season (14 May 2014), trajectories are continental and short-range. Air masses at low altitudes (500m, 1000m) appear to be originated from a western direction crossing Afghanistan and Pakistan following through Indian state Punjab and Haryana. Fig. (7a) and influenced by pollutants from nearby industrialized areas. In summer (13 June 2014), trajectories of air mass are marine type and occurred at heights (500m, 1000m) bring ozone and ozone precursors from southwestern directions i.e. Aarebic Sea and crossing some landmass area before reaching Uttarakhand. It is an indication of gasoline combustion originated from the Middle East and transport by the air mass trajectories at different altitudes fig (7b). Trajectories followed by air mass in the last week of June (end of summer) indicated the onset of monsoon from the Arabian Sea. Similarly, during the winter season (22 Dec 2014), trajectories are long-range and marine type transports from, originating from the Caspian Sea (northwestern direction of receptor site) and crossing Afghanistan Pakistan, and Indian states Punjab and also contributes to the observed law O3 levels due to oceanic influence fig (7c).

Since marine air mass is relatively clean compared to continental air mass. Therefore, this oceanic influence, the air mass increases the composition of hydroxyl radicals (OH). As ozone is a highly reactive and unstable gas so it reacts with OH radicals and converted into atomic oxygen or molecular oxygen. This causes depletion of O3 to a larger extent and maybe one of the reasons for the reduction of ozone concentration observed during the summer and winter season at this location. This complement our result illustrated in figure 4 i.e. maximum TCO value in spring then further decreases till winters. It concludes that maximum ozone concentrations in different seasons were a consequence of the transport effect of these trajectories.

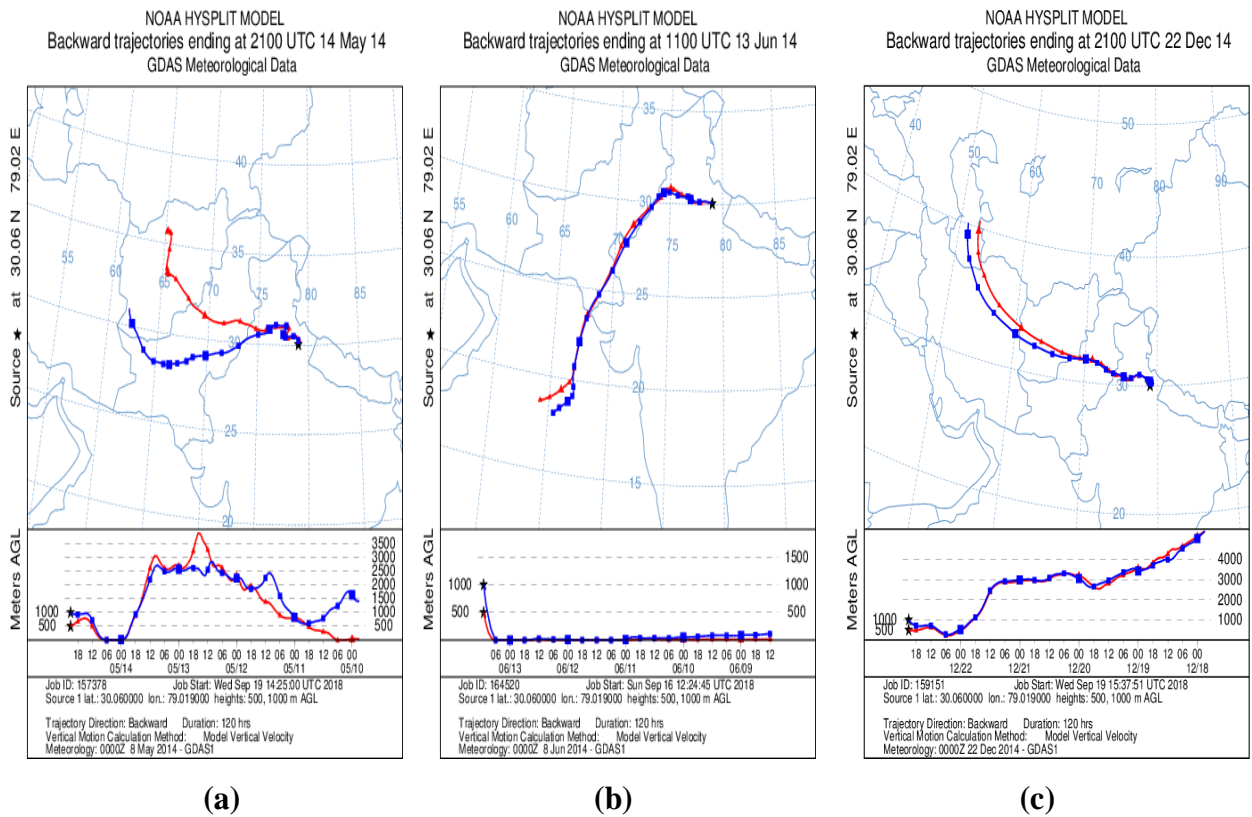

Figure 7: Back trajectories arriving in Uttarakhand (a) summer season (b) rainy season (c) winter season

# CONCLUSION

We have statistically analyzed spatial and temporal variability of TCO values over Uttarakhand (Environment of Himalaya) for 15 years i.e. 2005-2019. The daily data of TCO was retrieved from OMI and OMPS datasets having special coverage of 0.25×0.25 and 1×1 respectively. A significant seasonal and annual cycle has been established with the minimum value of TCO in December and maximum in May from both the datasets using daily TCO variability. The statistical outcomes for TCO from both the datasets are summarized. During the study period, we have obtained a monthly mean of TCO data retrieved from OMI was within ($297 \pm 13.78$) DU to ($260 \pm 8.73$) DU from OMI and ($300 \pm 16.08$) DU to ($258 \pm 6.07$) DU from OMPS respectively. Kolmogorov–Smirnov goodness of fit test was applied to find final distributions comply by the mean monthly series of TCO from both the instruments. We found that most of the months follow lognormal probability distribution at Uttarakhand for both the datasets of TCO.

Interannual variability of TCO has been performed for OMI datasets for 15 years and it showed merely the similar oscillating behavior for all six stations along the latitudinal and longitudinal belt of Uttarakhand. This fluctuation of TCO from 2005-2019 was an identical quasi-biennial oscillation (QBO) pattern. Further, we have observed an abrupt change in ozone variation in the year 2007 and 2016. This abnormal change might be due to interruption in quasi-biennial oscillation for the respective years.

Seasonal and annual TCO trend has been analyzed using the least square method and Mann-Kendall test for the seasonal and annual values of TCO from OMI datasets for 15 years over six selected sites of Uttarakhand. Corresponding results yield negative trends for annual, winters, and spring season with the maximum in winter while the positive trend for the rest of the seasons. Similar annual and seasonal trends have been estimated from statistical tests i.e. LSM and MK test. These results were highly correlated (0.86%) with each other. Although a negative annual trend has been observed in Haridwar, Pithoragarh, and US Nagar with a decreasing rate of -0.045DU/year, -0.057 DU/year, and -0.001 DU/year respectively but corresponding values are low which suggests a recovered TCO level in Uttarakhand from 2005 to 2019.

Furthermore, annual and seasonal COV pattern has been estimated in Uttarakhand. Annual COV of TCO showed a significant variation as a function of latitude which was increased from low to high latitude stations. The higher value of COV of ozone concentration occurred in winter whereas the lowest values occur at monsoon for all selected sites. Besides, seasonal variation was investigated through backward trajectories in association with continental and maritime during the study period. It showed that maximum ozone concentrations in different seasons are a consequence of the transportation of air mass trajectories. These results are footprints to correlate the TCO variation and climatic factors such as temperature, relative humidity, rainfall, and zonal wind, etc. Furthermore, we observed a positive skewed asymmetric curve of TCO for the higher latitude station of India Leh Laddak (34.10N-77.4E), Dehradun (28.37N-77.13E) and a negatively skewed curve for lower latitude station Banglore (12.58N-77.34E) and Kanyakumari (8.4N-77.32E). This was statistically confirming an earlier peak of maximum ozone concentration from northward to southward stations.

Apart from this, a comparative study for two data sets i.e. OMI and OMPS has been performed in terms of relative difference made between average monthly TCO data retrieved from both the datasets during the period of 8 years (January 2012 to December 2019) at Uttarakhand. The estimated average relative difference was merely 3% and in the range of 0.51% to 5.29%. Indeed, the daily data TCO from 2012-2019 was highly correlated for every year in the range of 68% to 99%. The outcomes of TCO distribution and trends were found similar to both the datasets. These results suggest that although both data sets used in this study i.e. OMI and OMPS have different resolutions but their measurements complement each other. Highly correlated outcomes from both these datasets enhance the validity of the results obtained in this study as well.

In summary, this study provides a detailed statistical interpretation and estimation of TCO variation over the Himalayan region i.e. Uttarakhand, India. Ozone layer depletion, mass trajectories, etc. have been analyzed and discussed using the data set of 15 years (2005-2019) retrieved from OMI and OMPS. The estimated results suggest a crucial variation in TCO has been observed in Uttarakhand over the study period. It is expected that the obtained outcomes would be helpful for the development of further studies of ozone variation (TCO variation) over other geographical regions as well.

## ACKNOWLEDGMENT

The authors express profound gratitude to (NASA) Goddard Space Flight Centre and NOAA national oceanic and atmospheric administration (NOAA) for making the OMI and OMPS data used for this work available on their web site. The authors gratefully acknowledge the NOAA Air Resources Laboratory (ARL) for the provision of the HYSPLIT transport Web site (http://ready.arl.noaa.gov) used in this publication.

## REFERENCES

A.Badawy, H. Abdel Basset, M. Eid, 2017. Spatial and Temporal Variations of Total Column Ozone over Egypt. Journal of Earth and Atmospheric Sciences 2, 1-16.

Anderson, J., J. M. Russell III, S. Solomon, and L. E. Deaver, Halogen Occultation, 2000. Experiment confirmation of stratospheric chlorine decreases in accordance with the Montreal Protocol. Journal of Geophysical Research 105, 4483–4490.

Ayodeji Oluleye, Emmanuel Chilekwu Okogbue, 2013. Analysis of temporal and spatial variability of total column ozone over West Africa using daily TOMS measurements. Atmospheric Pollution Research 4, 387-397.

Brasseur, G., and H. M., Hitchman, 1988. Stratospheric Response to Trace Gas Perturbations: Changes in Ozone and Temperature Distributions. Science 240, 634 637.

Brasseur, G., and Solomon, S. 1997. Aeronomy of the middle atmosphere, D. Reidel publishing company, New York, 2nd edition. 1–472.

Brune, W. H., D.W. Toohey, J. G. Anderson, and K. R. Chan, 1990. In situ observations of ClO in the Arctic stratosphere: ER-2 aircraft results from $59^oN$ to $80^oN$ latitude. Geophysical Research Letters 17, 505-508.

By Michael L. Stein, 2007. Spatial variation of Total Column Ozone on a global scale. The Annals of Applied Statistics 1, 191–210

Census, 2011. http://www.census2011.co.in/census,

Cicerone, R. J., 1974. Stratospheric ozone destruction by man-made chloroflouromethane. Science 185, 1165–1167.

Chakrabarty, D. K., and Chakrabarty, P., 1979. Ozone results from Indian stations using Dobson instruments, Proceedings of the NATO advanced institute on Atmospheric Ozone, 123-127

Chakrabarty, D.K., Peshin, S.K., Pandya, K.V., and Shah, N.C., 1998. Long-Term Trend of Ozone Column over the Indian Region. Journal of Geophysical Research 103, 19245-19251.

Crutzen, P. J., 1974. Estimates of possible future ozone reductions from continued use of fluoro-chloromethanes ($CF_2Cl_2$, $CFCl_3$). Geophysical Research Letters 1, 205–208.

David W. Fahey and Michaela I., 2014. Hegglin Coordinating Lead Authors Twenty Questions and Answers about the Ozone Layer: Update

Draxler, R. R., Rolph, G. D., 2011. HYSPLIT (HYbrid Single-Particle Lagrangian Integrated Trajectory), http://www.arl.noaa.gov/ready/hysplit4.html, NOAA Air Resources Laboratory.

Dobson, G. M. B., 1973. The laminated structure of the ozone in the atmosphere. Quarterly journal of the royal meteorological society 99, 599-607.

Dobson, G. M. B., 1968. Forty Years Research on Atmospheric Ozone at Oxford: a History. Applied optics 7, 387-405.

Farman, J. C., B. G. Gardiner, and J. D. Shanklin, 1985. Large Losses of Total Ozone in Antarctica Reveal Seasonal ClOx/NOx Interaction. Nature 315, 207-210.

Garcia, R., and Solomon, S., 1987. A Possible Relationship between Interannual Variability in Antarctic Ozone and Quasi-Biennial Oscillation. Geophysical Research Letters 14, 848-851.

Goutail.F, J.-P. Pommereau, C. Phillips, C. Deniel, A. Sarkissian, F. Lefèvre, E. Kyro, M. Rummukainen, P. Ericksen, S.B. Andersen, B.-A. Karastan-Hoiskar, G. Braathen, V. Dorokhov, V.U. Khattatov, 1999. Depletion of column ozone in the Arctic 1993–1994 and 1994–1995. Journal of atmospheric chemistry 32, 1-34.

Groves, K. S., and Tuck, A. F,1980. Stratospheric $O_3$-$CO_2$, coupling in a photochemical-radiative column model, 11: With chlorine chemistry, *Quart. J. R. Met. Soc., 106,* 141-157.

Guus J. M. Velders, Stephen O. Andersen, John S. Daniel, David W. Fahey, and Mack McFarland, 2007. The importance of the Montreal Protocol in Protecting Climate. National Academy of Sciences 104, 4814-4819.

Hema Bisht, Bimal Pande, Ramesh Chandra, Seema Pande, 2014. Statistical Study Solar activity Features with total column ozone at two hill stations of Uttarakhand. Indian Journal of Radio and Space Physics 43, 251-262.

Iwasaka, Y., Kondoh, 1987. Depletion of Antarctic ozone: Height of ozone loss region and its temporal changes. Geophysical research letters 14, 87-90.

J.I Freijer., J.C.H., Van Eijkeren, L.van Bree, 2002. A model for the effect on Health of repeated exposure to ozone. Environmental Modelling & Software 17, 553–562.

Kundu, N., and Jain, 1993. Total Ozone Trends over Low Latitude Indian Stations. Geophysical Research Letters 20, 2881-2883.

Kok Chooi Tan, Hwee San Lim, Mohd Zubir Mat Jafri, 2014. Analysis of total column ozone in Peninsular Malaysia retrieved from SCIAMACHY. Atmospheric Pollution Research 5, 42-51.

Madronich, S., McKenzie, R.L., Björn, L.O., and Caldwell, M. M., 1998. Changes in biologically active ultraviolet radiation reaching the Earth's surface. Journal of Photochemistry and Photobiology B: Biology 46, 5-19.

Mani, A., and Sreedharan, C.R., 1973. Studies of Variations in the Vertical Ozone Profiles over India. Pure and Applied Geophysics 108, 1180-1191.

Montreal protocol on substances that deplete the ozone layer, Final Act (1987), United Nations Environment Programme (UNEP 1987).

Nandita D. Ganguly and K.N.Iyer, 2005. Study of variations in Columnar Ozone Concentration at Rajkot. Journal of Indian Geophysical Union 9, 189-196.

Newman, P. A., Nash, E. R., Kawa, S. R., Montzka, S. A., and Schauffler, S. M., 2006. When will the Antarctic ozone hole recover?. Geophysical Research Letters 33, L12814.

Ningombam S S., 2011. Variability of sunspot cycle QBO and total ozone over high altitude western Himalaya region. Journal of Atmospheric and Solar-Terrestrial Physics 73, 2305-2313.

Ogunjobi K.O, Ajayi V.O, Balogun I.A., 2007. Long–term trend analysis of Tropospheric total column ozone in Africa. Research Journal of Applied Science 2, 280–284.

P. J. Nair, S. Godin-Beekmann, J. Kuttippurath, G. Ancellet, F. Goutail, A. Pazmiño, L. Froidevaux, J. M. Zawodny, R. D. Evans, H. J. Wang, J. Anderson, and M. Pastel, 2013. Ozone trends derived from the total column and vertical profiles at a northern mid-latitude station. Atmospheric Chemistry and Physics 13, 10373-10384.

Pulikesi, M., Baskaralingam, P., Rayudu, V.N., Elango, D., Ramamurthi, V., Sivanesan, S., 2006. Surface ozone measurements at urban coastal site Chennai. India. Journal of Hazardous Materials 137, 1554–1559.

P. K. Bhartia, R. D. McPeters, L. E. Flynn, S. Taylor, N. A. Kramarova, S. Frith, B. Fisher, and M. DeLand, 2013. Atmospheric Measurement Techniques Solar Backscatter UV (SBUV) total ozone and profile algorithm. Atmospheric Measurement Techniques 6, 2533–2548.

Pinedo-Vega, J.L., Molina-Almaraz, M., Ríos-Martínez, C., Mireles-García, F. and Dávila-Rangel, J.I. 2017. Global and Hemispherical Interannual Variation of Total Column Ozone from TOMS and OMI Data. Atmospheric and Climate Sciences, 7, 247-255.

Rowland F. S., 1989. Chlorofluorocarbons and the depletion of stratospheric ozone. American scientist 77, 36-45.

Rowland F. S., 1991. Stratospheric Ozone Depletion. Annual Review of Physical Chemistry 42, 731-768.

R. Venkanna, G. N. Nikhil, T. Siva Rao, 2015. Environmental monitoring of surface ozone and other trace gases over different time scales: chemistry, transport, and modeling. International Journal of Environmental Science and Technology 12, 1749–1758.

Rubin, M. B., 2001. The History of ozone. The Schönbein Period, 1839-1868. Bulletin for the History of Chemistry 26.

Schmalwieser, A. W., Schauberger, G., and Janouch, M., 2003. Temporal and spatial variability of total ozone content over central Europe: Analysis in respect to the biological effect on plants. Agriculture and Forest Meteorology 120, 9-26.

Stolarski, R. S., Schoeberl, M. R., Newman, P. A., McPeters, R. D., and Krueger, A. J., 1990. The 1989 Antarctic ozone hole as observed by TOMS. Geophysical Research Letters 17, 1267–1270.

Sivasakthivel.T and K.K.Siva Kumar Reddy, 2011. Ozone Layer Depletion and Its Effects: A Review. International Journal of Environmental Science and Development 2.

Stohl, A., 1998. Computation, accuracy and applications of trajectories – a review and bibliography. Atmospheric Environment 32, 947–966.

Solomon, S., 1999. Stratospheric ozone depletion: a review of concepts and history. Review of Geophysics 37, 275–316.

UNEP, 1998. Environmental Effects of Ozone Depletion. United Nations Environment Programme

UNEP, 2003. Environmental Effects of Ozone. Depletion: 2002 Assessment. Photochemistry and Photobiology, 2:1–72. United Nations Environment Programme.

UNEP, 2016. Environmental Effects of Ozone. Depletion and its interactions with climate change: progress report, 2016 Assessment. Photochemistry and Photobiology, 2: 107-145.2017 United Nations Environment Programme.

Vazhathottathil Madhu, 2014. Spatial and Temporal Variability of Total Column Ozone over the Indian Subcontinent: A Study Based on Nimbus-7 TOMS Satellite. Atmospheric and Climate Sciences 4, 884-898.

Wayne, R. P., 1987. The Photochemistry of Ozone, Atmospheric Environment 21, 1683-1694.

Wayne, R. P., 2000. Chemistry of Atmospheres, chap. Ozone in Earth's stratosphere, 155-320, 3rd ed., Oxford University Press Oxford,

Weber, M., Dikty, S., Burrows, J. P., Garny, H., Dameris, M., Kubin, A., Abalichin, J., and Langematz, U., 2011. The Brewer-Dobson circulation and total ozone from seasonal to decadal time scales, Atmospheric Chemistry and Physics 11, 11221–11235.

Willett, H.C., 1962. The Relationship of Total Atmospheric Ozone to the Sunspot Cycle. Journal of Geophysical Research 67, 661-670.

World Meteorological Organization, 1989, Scientific Assessment of Stratospheric Ozone, WMO Report No.20.

World Meteorological Organization (WMO). 2011. 'Scientific assessment of ozone depletion: 2010', Global Ozone Researchand Monitoring Project Report No. 52. WMO: Geneva, Switzerland

Zubov, 2001. Assessment of the Effect of the Montreal Protocol on Atmospheric Ozone. Geophysical Research Letters 28, 2389–2392